\documentclass{cupconf}
\usepackage{graphicx}

  \checkfont{eurm10}
  \iffontfound
    \IfFileExists{upmath.sty}
      {\typeout{^^JFound AMS Euler Roman fonts on the system,
                   using the 'upmath' package.^^J}%
       \usepackage{upmath}}
      {\typeout{^^JFound AMS Euler Roman fonts on the system, but you
                   dont seem to have the}%
       \typeout{'upmath' package installed. cupconf.cls can take advantage
                 of these fonts,^^Jif you use 'upmath' package.^^J}%
      }
  \else
  \fi

  \checkfont{msam10}
  \iffontfound
    \IfFileExists{amssymb.sty}
      {\typeout{^^JFound AMS Symbol fonts on the system, using the
                'amssymb' package.^^J}%
       \usepackage{amssymb}%
         
       \let\ge=\geqslant  
      }{}
  \fi

  \IfFileExists{amsbsy.sty}
    {\typeout{^^JFound the 'amsbsy' package on the system, using it.^^J}%
     \usepackage{amsbsy}}
    {}

\newsavebox{\astrutbox}
\sbox{\astrutbox}{\rule[-5pt]{0pt}{20pt}}

\newcommand\hb{H${\beta}$~}
\newcommand\ha{H${\alpha}$~}
\newcommand\nii{[N{\sc ii}]~}
\newcommand\civ{[C{\sc iv}]~}
\newcommand\ciii{[C{\sc iii}]~}

\newcommand\siii{[S{\sc iii}]~}

\newcommand\oi{[O{\sc i}]~}

\newcommand\oiii{[O{\sc iii}]~}

\newcommand\sm{M$_{\odot}$}

\title[Magellanic planetary nebulae]{Magellanic Cloud planetary
nebulae as probes of stellar evolution and populations}

\author[L. Stanghellini]%
{L\ls E\ls T\ls I\ls Z\ls I\ls A\ns S\ls T\ls A\ls N\ls G\ls H\ls E\ls L\ls L\ls I\ls N\ls I\ls
}
\affiliation{Space Telescope Science Institute, and European Space Agency}

\begin{document}

\maketitle

\begin{abstract}

Planetary Nebulae (PNs) in the Magellanic Clouds offer the unique 
opportunity to study both the population and evolution of low- and 
intermediate-mass stars, in an environment that is free of the 
distance scale bias and the differential reddening that hinder the
observations of the Galactic sample. The study of LMC and SMC PNs also
offers the direct comparison of stellar populations with different
metallicity. The relative proximity of the Magellanic Clouds allows
detailed spectroscopic analysis of the PNs therein, while the 
{\it Hubble Space Telescope (HST)} is necessary to 
obtain their spatially-resolved images.
In this paper we discuss the history and evolution of this relatively 
recent branch of stellar astrophysics by reviewing the pioneering
studies, and the most recent ground- and space-based achievements.
In particular, we present 
the results from our recent {\it HST}
surveys, including the metallicity dependence of PN identification (and,
ultimately, the metallicity dependence of PN counts in galaxies);
the morphological analysis of Magellanic PNs, and the correlations between 
morphology and other nebular properties; the relations between
morphology and progenitor mass and age; and the direct analysis of
Magellanic central stars and their importance to stellar
evolution.
Our morphological results are broadly 
consistent with the predictions of stellar evolution if the progenitors 
of asymmetric PNs have on average larger masses than the progenitors of 
symmetric PNs, without any assumption or relation to binarity of the
stellar progenitors. 
\end{abstract}

\section{Introduction}

Planetary Nebulae (PNs) are the gaseous relics of the envelopes ejected by
low- and intermediate-mass stars
(1$<M<$8 \sm) at the tip of the asymptotic giant branch (AGB), thus 
they are important probes of stellar evolution, stellar
populations, and cosmic recycling. PNs have been observed
in the Local Group (as well as in external galaxies), probing stellar evolution and 
populations in relation to their environment.

The details of the observations of Galactic PNs and their central stars (CSs)
typically surpass the 
details of stellar and hydrodynamic models. Galactic PN studies are a
necessary background
toward the understanding the PN population in Local Group galaxies. 
Yet, the distance scale of Galactic PNs is uncertain to such a degree that 
the meaning of the comparison between observations and theory is hindered. 
By the same token, statistical studies 
of PN populations in the Galaxy suffer for the observational
bias against the detection of Galactic disk PNs, and for the patchy interstellar
extinction. 

PNs in the Magellanic Clouds (LMC, SMC), hundreds of low-extinction 
planetaries at uniformly known distances, are a real bounty for the stellar
evolution scientist. The composition gradient between the LMC, the SMC,
and the Galaxy, afford the study of the effects
of environment metallicity on PN evolution. The relative vicinity of the Clouds,
and the spatial resolution that can be achieved with the {\it Hubble Space Telescope (HST)}, 
allow the detection of PN morphology. 
Studying the PNs in the
Magellanic Clouds is a perfect example of how the Local Group 
can be efficiently (and uniquely) used as an astrophysical laboratory.
In this paper we review the history and the evolution of
this field of study, with particular focus on the results from
our recent {\it HST} Magellanic PN programs.

\section{Pioneers in the field, and recent developments}

PN studies in the Magellanic Clouds are relatively recent. 
The first detection and spectral identification of Magellanic PNs is due to
Lindsay (1955). Studies of Magellanic PN samples became 
common, and their importance evident, in the early sixties
(Aller 1961; Westerlund 1964). 
Ground based observations of Magellanic PNs suffer from the fact that
the contributions of nebular and stellar radiation are 
superimposed. Attempts to measure
the CS magnitudes have been hampered by the difficulties in separating
stellar and nebular contributions (e.g., Webster 1969).

Observations with the IUE, combined with the optical spectra acquired from the 
ground, have allowed the abundance analysis of Magellanic PNs. 
Space observations in the UV range were used for
the detection of the complete set of carbon lines at various ionization stages, 
and made the carbon abundance derivation much more reliable. 
The key results in abundance studies can be found, to name 
a few, in Peimbert (1984), Boroson \& Liebert (1989), Kaler \& Jacoby 
(1991). Optical spectroscopy of large samples of Magellanic PNs
have been carried out by Dopita and collaborators (Meatheringham \& Dopita 
1991ab; Vassiliadis, Dopita, Morgan, \& Bell 1992).
IUE observations were also used to measure the stellar luminosity beyond
the Lyman limit for the CSs, giving an estimate of the total stellar
luminosity, and and approximate estimate to the mass (Aller et al. 1987).
Several papers on Magellanic PN spectroscopy, 
abundances, and the connection of nebular and stellar evolution can be
found in the proceedings of IAU Symposia 
on planetary nebulae (e.g., Westerlund 1968; Feast 1968; 
Webster 1978), while the most recent, complete review on Magellanic PNs
is due to Barlow (1989). 

Studies of the chemical content of Magellanic PNs 
have been active in the recent past as well.
On the observational side, Leisy \& Dennefeld (1996), and Costa, de Freitas Pacheco,
\& Idiart (2000), have estimated new chemical 
abundances 
for several Magellanic PNs from optical and UV observations, enriching the databases
for studies on the dredge-up of post-AGB stars and on the ISM enrichment in galaxies.
On the theoretical side, van 
den Hoek \& Groenewegen (1997) calculated new chemical yields of the 
interstellar medium enrichment from synthetic evolution of intermediate-mass 
stars. With models from a wide range of initial masses and metallicities (included those
appropriate for the Magellanic populations), van 
den Hoek \& Groenewegen confirm that the yields of nitrogen and carbon 
change abruptly for M$>$3 \sm, due to the hot-bottom burning effect. 

To date, 277 LMC PNs (Leisy, Dennefeld, Alard, \& Guibert 1997) and 55 SMC PNs (Meyssonnier \& Azzoppardi
1993) are known. The total number of Magellanic PNs have more than doubled
from the last count by Barlow (1989).
In the last few years, several emission line surveys have been completed,
or are near completion (e.g., UKST survey:
Morgan 1998, Parker \& Phillips 1998; UM/CTIO survey: Smith et al. 1996). 
Future analysis of these surveys is
essential for the future health of Magellanic PN research. We expect that the PN counts 
in the Clouds will increase significantly, improving the statistical significance
of these studies.
One important aspect of these surveys is the
discovery of fainter PNs, that contributes to
increasing the reliability of the faint end of the Magellanic PN 
luminosity function, and to enlarge the pool of known evolved PNs.

Related to the populations of Magellanic PNs is the 2MASS survey (Egan, Van Dyk, \& Price 2001). 
The importance of this multi-wavelength infrared survey to LMC PNs is
related to the spatial distribution of the different types of AGB stars. 
Egan et al. showed that low mass AGB stars occupy the whole of the 
LMC projected volume, while the higher mass, younger population, AGB stars populate 
preferentially the LMC bar. Chemical and morphological studies of large LMC PN samples 
should be compared to the AGB samples, to relate the PN populations in the 
LMC to their immediate evolutionary progenitors.

\section{Early Hubble Space Telescope observations}

Extended studies of Galactic PNs have shown that PN morphology 
is intimately related to the mass and evolution of their CSs,
to their stellar progenitors, and to the nebular chemistry. 
In the case of the LMC and
the SMC, morphological studies became possible with the use of the
cameras on the {\it HST}. 
The {\it HST} has also the capability of spatially
separate the image of the nebula and that of the CS,
making direct stellar analysis possible.

The early narrow-band images of 
Magellanic PNs were obtained before the first {\it HST} servicing mission 
(i.e., before the 
installation of COSTAR on {\it HST}) with the {\it Faint Object Camera} 
(Blades et al. 1992). 
Other images by Blades et al. have later been published by 
Stanghellini et al. (1999),
where the quality of the pre-COSTAR images was validate through 
their comparison with post-COSTAR images of the same objects. 
These papers have made available
15 Magellanic PN images usable for statistical and morphological studies, while another 15 LMC 
PNs have been observed with the {\it Planetary Camera 1} by Dopita et al. (Dopita et al. 1996; 
Vassiliadis et al. 1998). Finally, an additional ten {\it Wide Field and Planetary Camera 2}
narrow-band images of LMC PNs are available in the 
Hubble Data Archive (program 6407).

\section{Our Magellanic PN program}

During the {\it HST} Cycle 8 we started a series of surveys aimed at obtain the 
size, morphology, and CS properties of all Magellanic PNs known to date.
The {\it HST} was an obligatory choice, since the Magellanic PNs are typically
half an arcsec across, thus they are generally not resolved with ground-based
telescopes. 

The medium-dispersion, slitless capability of STIS offers us a valuable 
opportunity to study the evolution and morphology of the Magellanic
Cloud PNs and 
their CSs at once.  We have applied this capability in several
SNAPSHOT surveys, obtaining images in the light of up to 7 of 
the most prominent low- and moderate-ionization optical, nebular emission 
lines.  We also obtained direct continuum images to identify the 
correct CS (in spite of the crowded fields), and 
to measure the optical continuum emission.

\begin{table}
\label{tab1}  
\caption{STIS observations of Magellanic PNs}
\begin{center}
\begin{tabular}{lllll}
&&&&\\
Program & Galaxy& PNs& Observing mode& Papers$^a$\\
&&&&\\  
8271&    LMC& 29& G750M/G430M/50CCD& I, II\\
8663&	   SMC& 29$^b$& G750M/G430M/50CCD& III, IV\\
9077&    LMC& 52&  G750M/G430M/50CCD& V\\
9120&    LMC& 28& G140L/G230L&  VI\\
    \end{tabular}

$^a$ Papers V and VI are in preparation; 
$^b$ Two PNs in this sample are misclassified H II regions.
  \end{center}
\end{table}

In addition to the optical slitless spectra and broad band continuum images
of the LMC and SMC PNs, we have acquired STIS UV spectra of 24 LMC PNs.
In the cases where the CSs were hard to find in our
STIS broad band images, we have also
acquired WFPC2 Str\"omgren images (Program 8702).
We have used through our investigations the limited data available
in the literature (see Stanghellini et al. 1999). 
In Table 1 we list the major {\it HST} data sets of 
our Magellanic PN project. The
actual data can be easily retrieved from the dedicated MAST Archival page
{\it http://archive.stsci.edu/hst/mcpn}. In addition to the {\it HST} data we
have made extensive use of the spectra acquired from the ground by
us (papers in preparation), and available in the literature.

In Figure 1 we show a sampler of the most common morphological types
of Magellanic PNs, from the broad-band images in our surveys. These PNs are 
more than 50 times farther away than the typical galactic PNs, yet the major 
morphological features are easily recognized, as are the location of their
CSs, when visible. 

In the following sections we describe some of the most
important results that we obtain from the analysis of our PN samples.

\begin{figure}
\centering
 \includegraphics[width=3.8cm]{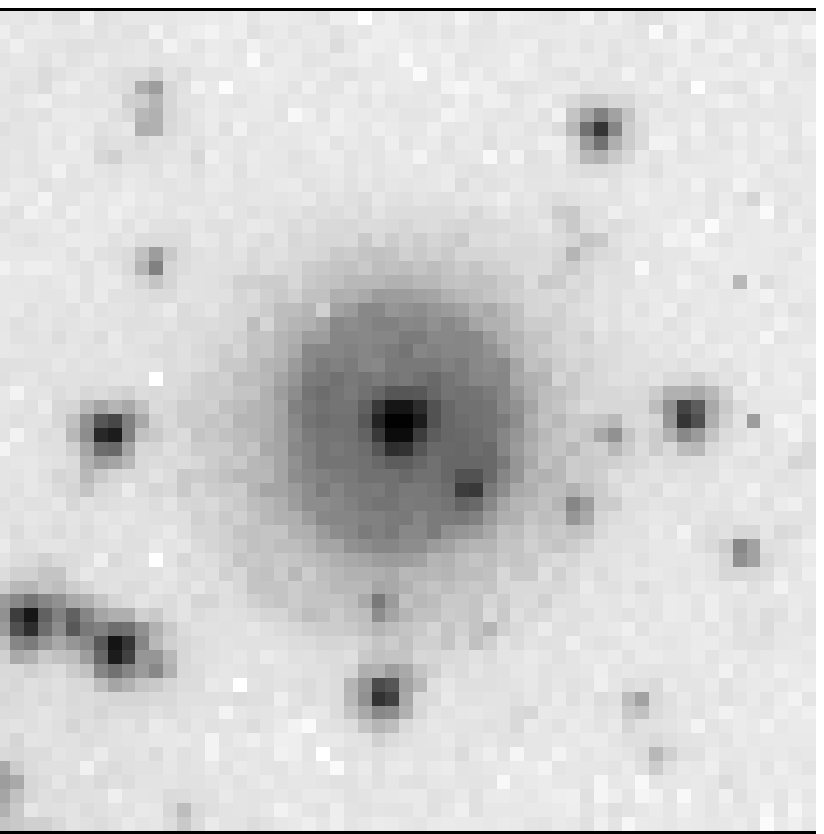}%
 \hspace{1cm}%
 \includegraphics[width=3.8cm]{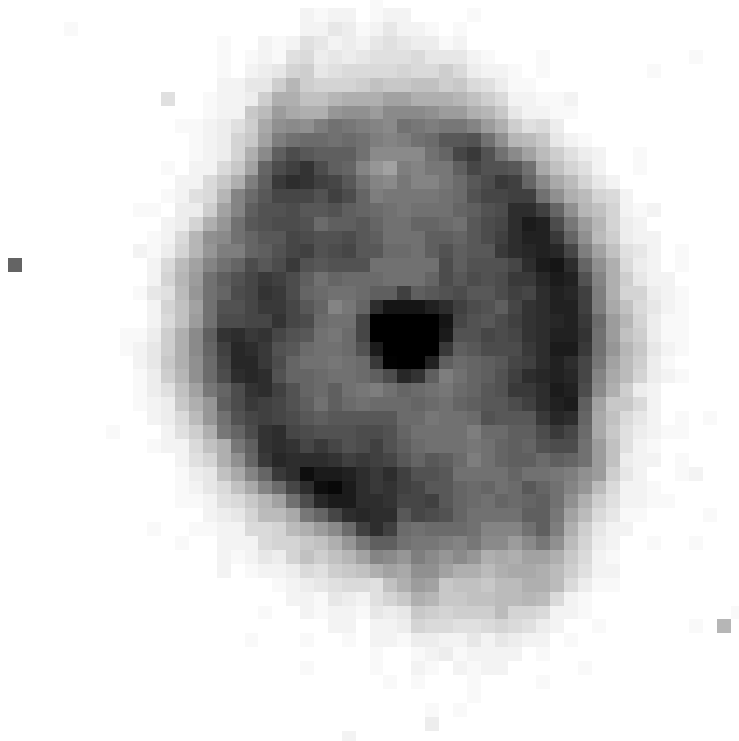}%
\hspace{1cm}%
 \includegraphics[width=3.8cm]{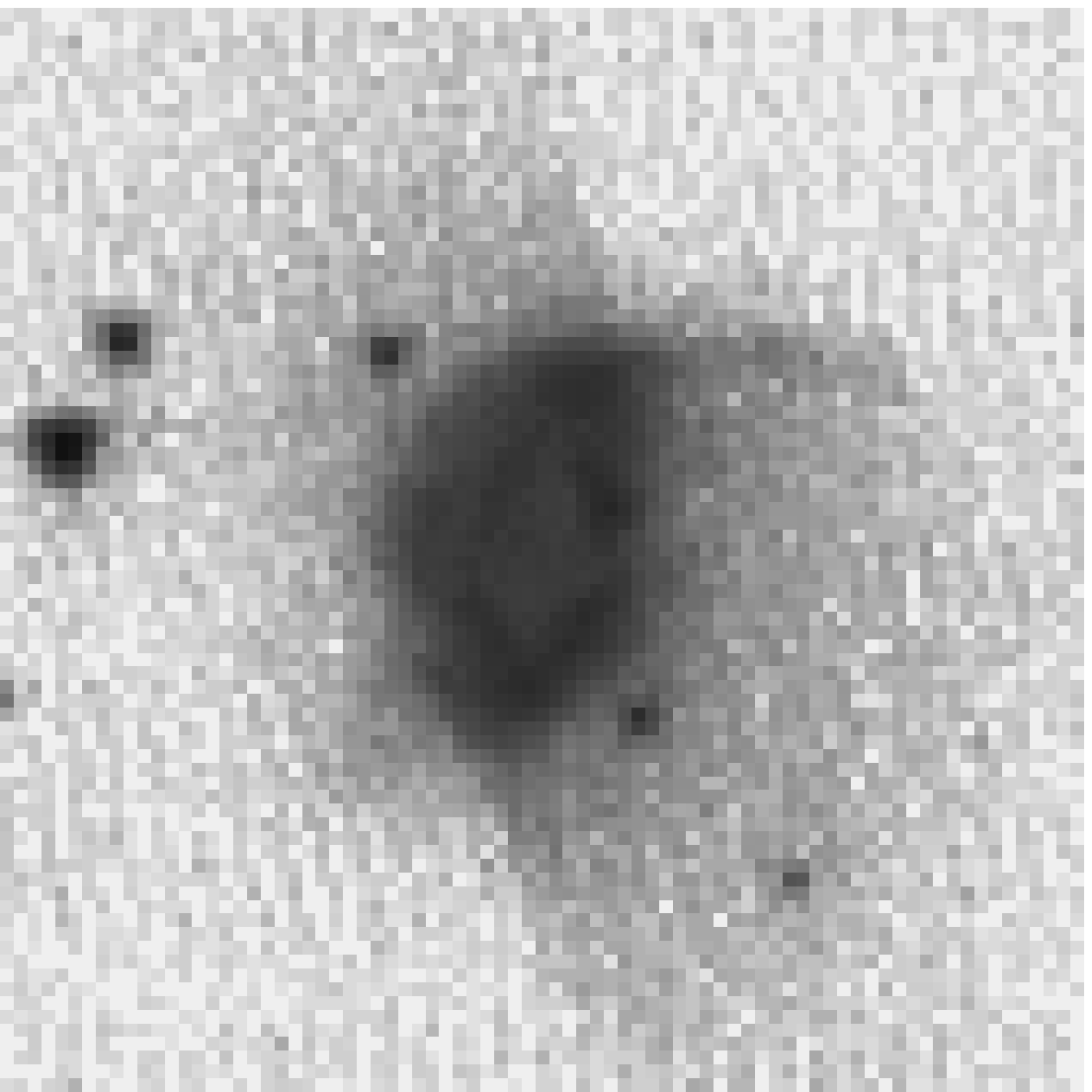}%
\hspace{1cm}%
 \includegraphics[width=3.8cm]{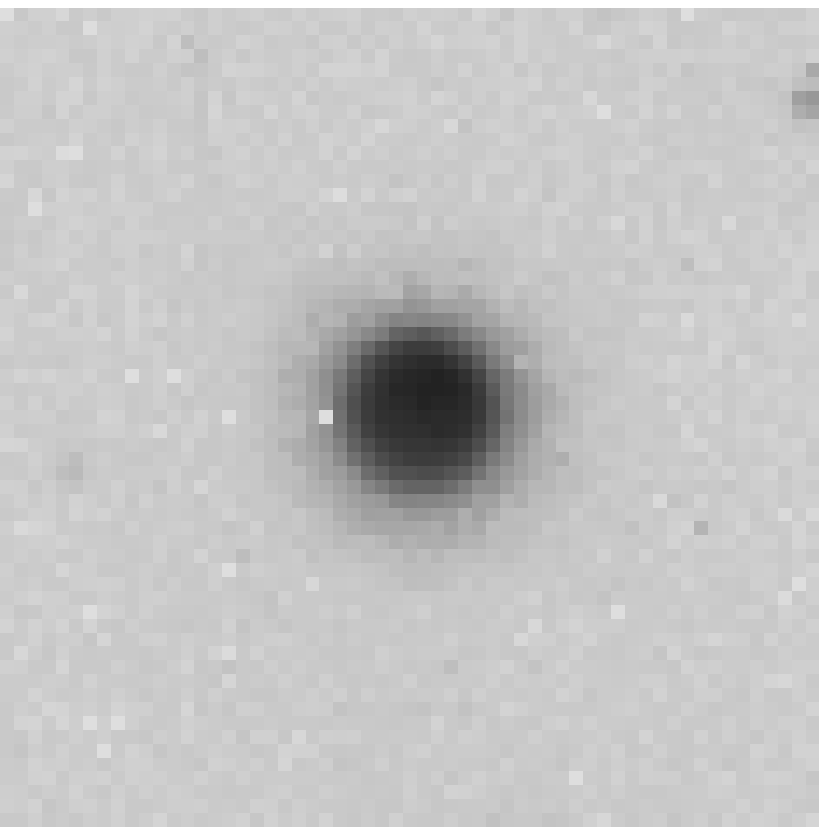}%
\hspace{1cm}%
 \includegraphics[width=3.8cm]{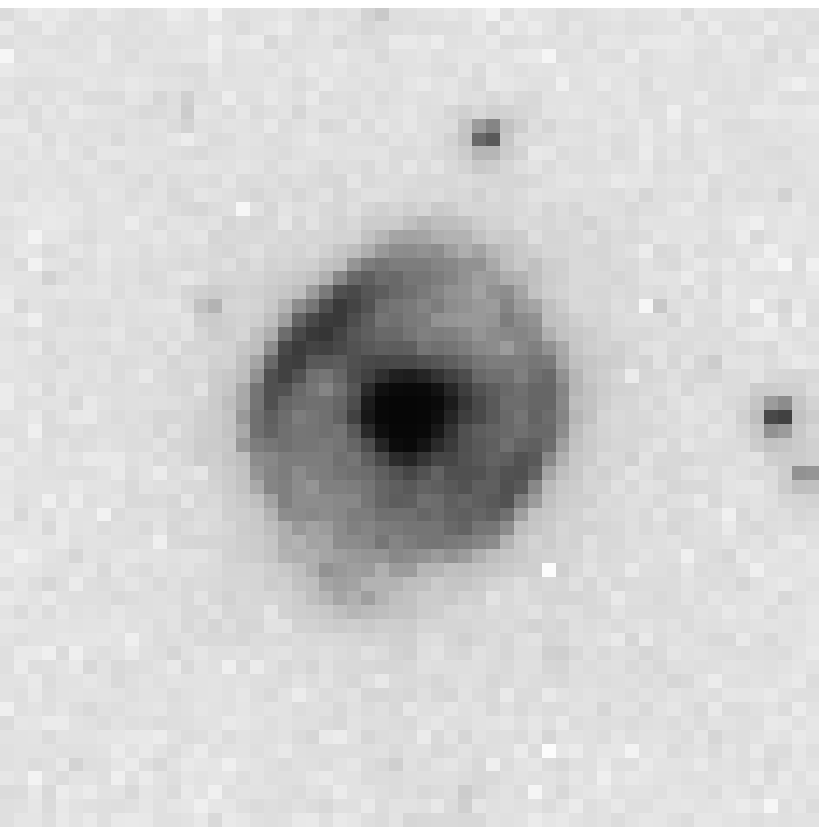}%
 \hspace{1cm}%
 \includegraphics[width=3.8cm]{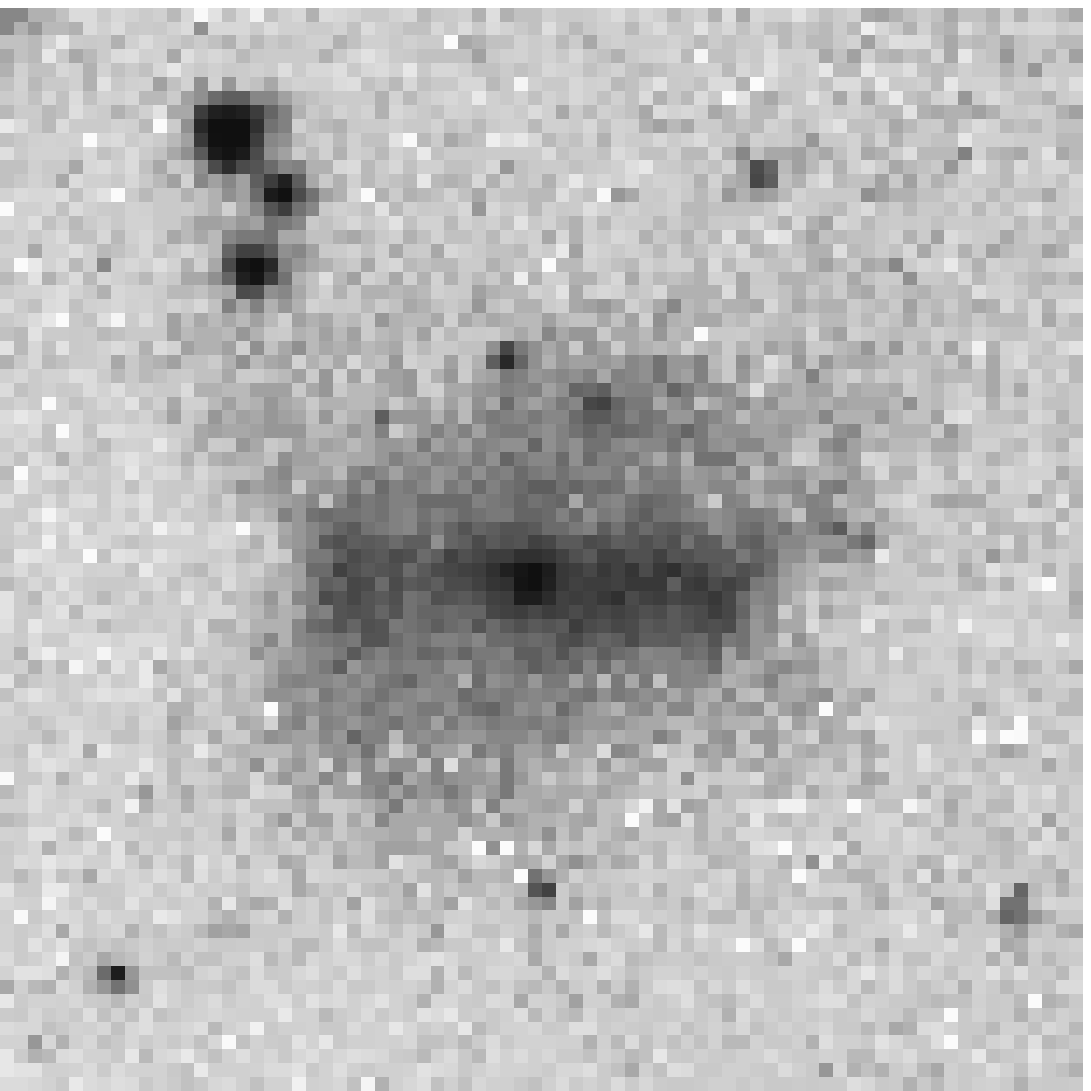}%
  \hspace{1cm}%
 \includegraphics[width=3.8cm]{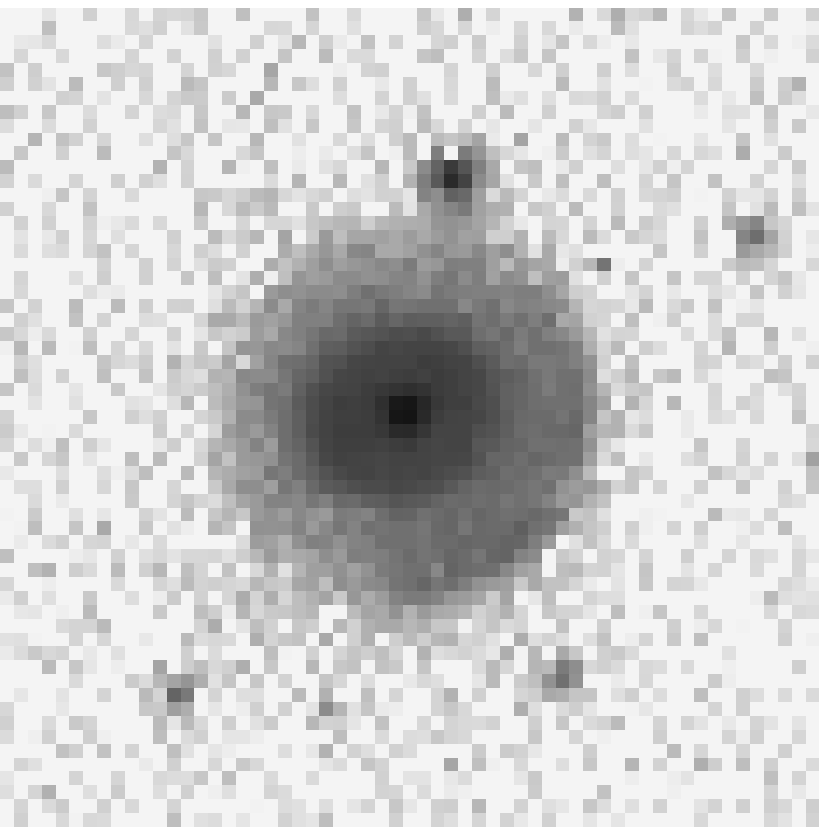}%
 \hspace{1cm}%
 \includegraphics[width=3.8cm]{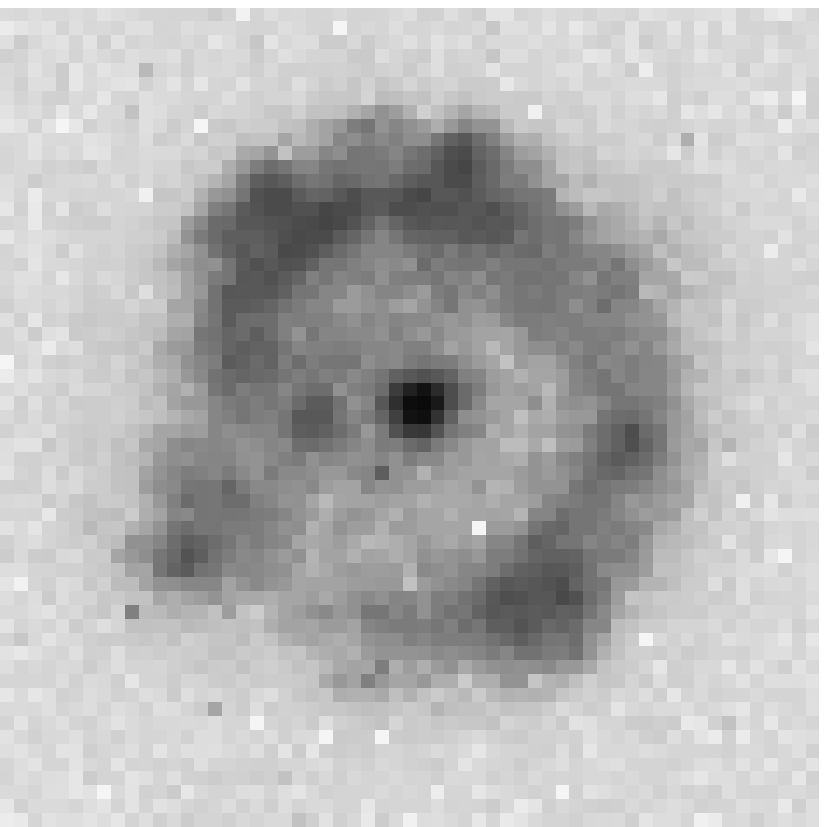}%
 \hspace{1cm}%
 \includegraphics[width=3.8cm]{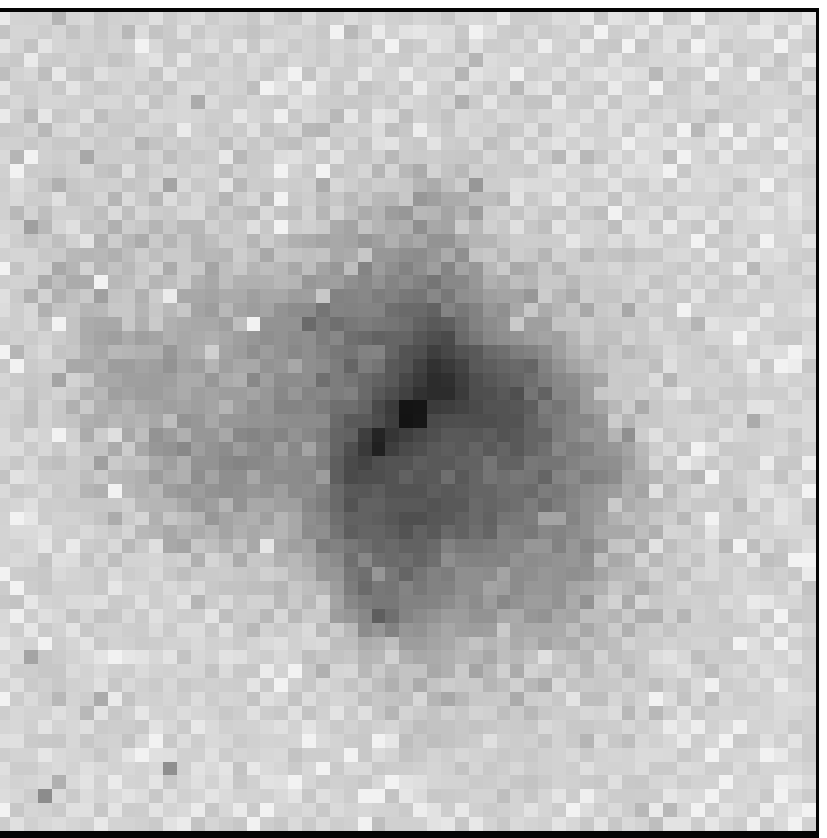}%

    \caption{Magellanic PNs. Left panel: round PNs LMCJ~33, 
    SMC-SP~34, and LMC-MG~40; Central panel: elliptical PNs LMC-SMP~101, SMC-MG~8,
    and SMC-MG~13; Right panel: bipolar PNs  LMC-SMP~91, SMC-MA~1682,
    and LMC-MG~16. All thumbnails are 9 arcsec$^2$ sections of the
    STIS broad-band images. 
    }\label{fig1}
\end{figure}

\subsection{Misclassified planetary nebulae}

Extra-Galactic PNs in general, and Magellanic PNs in particular,
are typically discovered with \ha or \oiii~$\lambda$5007 
surveys, with the on- and off-band
observations technique. 
The PN nature of the \oiii $\lambda$5007-bright object can be
generally confirmed by means of
spectroscopic follow-up. Marginal misclassification is possible
even after spectroscopic analysis, and
only the spatially
resolved observations can ultimately confirm the PN nature.

Among the ~230 LMC PNs known, about 80  have been observed with {\it HST} 
by us, and another 30 had been observed earlier by Dopita and Blades.
In the SMC, we acquired {\it HST} observations of 29 targets, of the 55 known
SMC PNs. The spatially resolved sample consists of the \oiii $\lambda$5007
bright half of the 
known Magellanic PNs, given that we had selected the brightest PNs in order to fit
the observations within one {\it HST} orbit. Thus our sample is representative
of the (bright) Magellanic PNs. 

It is then extremely interesting to note that we confirmed 100$\%$ of the
LMC targets to be PNs, while 10$\%$ of the SMC targets were, instead, 
ultra-compact H II regions around very young star clusters
(Stanghellini, Villaver, Shaw, \& Mutchler 2003b). This result bears on the 
validity of the PN luminosity function (PNLF) as a secondary
distance scale indicator. In fact, the extra-Galactic PNLF is populated
by the \oiii bright PNs of each studied population, and our result seems
to indicate that such populations may be contaminated by H II regions 
in different ways depending on the galaxian metallicity.

\subsection{PN morphology}

The morphology of Galactic PNs has been studied rather thoroughly in the past
decade, and it has been found that the morphological types correlate with
the PN progenitor's evolutionary history, and the stellar mass. There is strong evidence
that asymmetric (e. g., bipolar) PNs are the progeny of the massive
AGB progenitors (3-8 \sm). Bipolar PNs are nitrogen enriched and carbon poor
(Stanghellini, Villaver, Manchado, \& Guerrero 2002b). The analysis of the morphological 
types and their distribution in a PN population is then very useful to infer
the age and history of a given stellar sample.

Galactic PNs have been classified as round, elliptical, bipolar (and quadrupolar), bipolar
core (those bipolar PNs whose lobes are too faint to be detected, but whose 
equatorial ring is very evident), and point-symmetric.
The majority of Galactic PNs are elliptical, but the actual number of
bipolars could be underestimated,
given that they typically lie in the Galactic plane (i.e., they may suffer high 
reddening).
PNs in the Clouds, when spatially resolved, show the same admixture of morphological types 
than the Galactic PNs (see Fig.~1). While we do not attempt a statistical comparison of the
MC and Galactic PN morphological types, given the selection effects that hamper
Galactic PNs, we can meaningfully compare the LMC and SMC samples. Both
samples suffer from low field extinction, and they have been preselected in
more
or less the same way.

\begin{table}
\label{tab2}  
\caption{Morphological distribution}
\begin{center}
\begin{tabular}{lll}
&&\\
Morphological type& $\%$ LMC&  $\%$ SMC\\
&&\\  
Round (R)&		29&	35\\
Elliptical (E)&		17&	29\\
R+E (symm.)&		46&	64\\
Bipolar (B)&		34&	6\\
Bipolar core (BC)&	17&	24\\
B+BC (asymm.)&		51&	30\\
Point-symmetric (P)&	3&	6\\	

    \end{tabular}

\end{center}
\end{table}

The results of the morphological distribution in the Clouds is summarized in Table 2.
Together with the percentage in each morphological class, we give
the total of symmetric (round and elliptical) and asymmetric (bipolar and bipolar
core) PNs. One striking difference between the two distributions is that the fraction of
bipolar PNs in the LMC is almost six times that of the SMC. Bipolar PNs are easily 
recognized, thus this is a sound result. If we add to the asymmetric PN count the 
bipolar core PNs, we obtain that half of the LMC PNs are asymmetric, while only
a third of the SMC PNs are asymmetric. Observational biases play in the same way for the
two samples.

What insight can we get from the morphological results? First of all, it is
clear that the set of processes that are involved in the formation of the
different PN shapes are at work {\it in all galaxies where morphology has been
studied}. Second, the SMC environment may disfavor the onset
of bipolarity in PNs. Otherwise, the different morphological statistics 
may indicate different populations of stellar progenitors in the two Clouds.
While it seems reasonable to conclude that a low metallicity environment 
is unfavorable to bipolar evolution, the exact causes have not been studied
yet. A detailed study of metallicity and mass loss may clarify this 
point. On the other hand, the different morphological statistics may simply 
be related to a lower average stellar mass of the PN progenitors in the SMC.
If this was the case, we should observe also lower CS masses in the
SMC PNs than in the LMC PNs. Our preliminary measurements seem to indicate that
this is also the case (Villaver, Stanghellini, \& Shaw, in preparation).

\subsection{Surface brightness evolution}

\begin{figure}
 \includegraphics[height=12truecm]{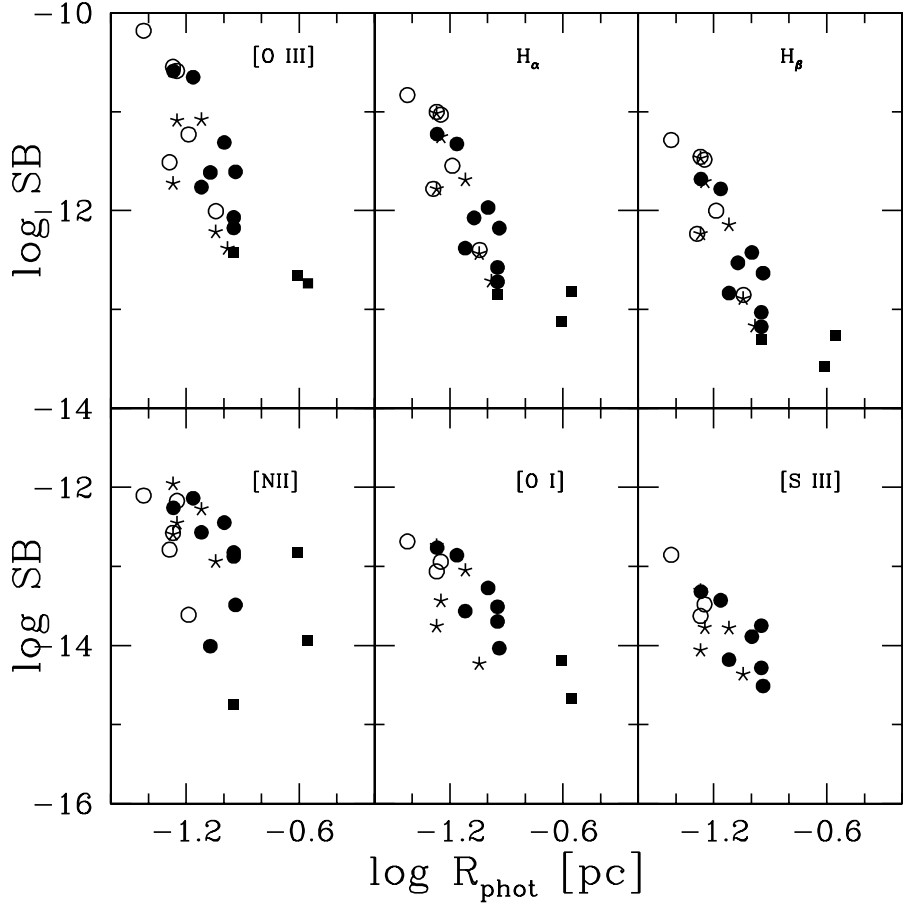}
 \caption{Surface brightness decline from the {\it HST} multiwavelength images of the LMC
  and SMC PNs (adapted from Stanghellini et al. 2002a). 
Emission lines in which the SB is derived are indicated in the panels (see also text). 
Symbols indicate morphological types: round 
({\it open circles}), elliptical ({\it asterisks}), bipolar core ({\it filled circles}), and 
bipolar and quadrupolar ({\it filled squares}). The photometric radii are measured from the 
\oiii $\lambda$5007 images, where available.}\label{fig2}
\end{figure}

The surface brightness of LMC and SMC PNs correlates with the photometric
radius, as illustrated in Figure 2. In the figure we plot the logarithmic
surface brightness, derived from the observed total line fluxes and 
the apparent radii, versus the photometric radii of the nebulae, measured as
the distance from the CS (or the geometrical center of the nebulae)
where the enclosed flux is 85$\%$ of the total nebular flux (see
Stanghellini et al. 1999). The six panels illustrate the surface brightness-
radius relation for the major six emission lines: \oiii $\lambda$5007,
\ha, \hb, \nii $\lambda$6584, \oi $\lambda$6300, and \siii$\lambda$6312. 

The surface brightness-
photometric radius relation is tight in all spectral lines, with the 
exception of the \nii emission line, where a larger spread is present,
particularly for bipolar PNs. A possible factor is the
larger range of nitrogen abundances in bipolar and BC PNs. 
The surface brightness-
photometric radius relations hold
only in the cases in which the nebular density $N_{\rm e}$ is smaller
than the  critical density, $N_{\rm crit}$ (the density at which the
collisional de-excitation rate balances the radiative  transition rate).

A good eye-fit to the surface brightness-
photometric radius relation is SB $\propto$ R$_{\rm phot}^{-3}$. This relation
can be reproduced via hydrodynamic modeling of evolving PNs and their
CSs (Villaver \& Stanghellini, in preparation). The surface brightness-
photometric radius relation in the light of \ha (or \hb) is tight enough that
it can be used to set the distance scale for Galactic PNs with intrinsic
uncertainties of the order of 30$\%$ or less (Stanghellini et al. in preparation), 
while the current calibration of the Galactic PN distance scales carry errors
of the order of 50$\%$ or more.

In Figure 2 we note that the symmetric (round and elliptical) PNs tend to
cluster at high surface brightness and low radii, while the asymmetric
PNs occupy the lower right parts of the diagrams. This separation can be
interpreted with a slower evolutionary rate for the symmetric PNs, 
which agrees with the idea that symmetric PNs derive from lower mass progenitors
(Stanghellini, Corradi, \& Schwarz 1993, Stanghellini et al. 2002b). 

\subsection{The \oiii/\hb distribution}

In Figure 3 we plot a histogram of the ratio of (reddening-corrected)
fluxes of the \oiii $\lambda$5007 and \hb lines 
for the PNs of the SMC and the LMC.  The median of the 
SMC distribution is a factor of two lower than for the corresponding LMC 
distribution (
$<$\oiii/\hb$>_{\rm SMC}$ = 5.7 $\pm$ 2.5 and $<$\oiii/\hb$>_{\rm LMC}$ = 9.4 $\pm$ 3.1).  
This result is free of object selection 
biases since both sets of targets were chosen in much the same way.

\begin{figure}
 \includegraphics[height=8truecm]{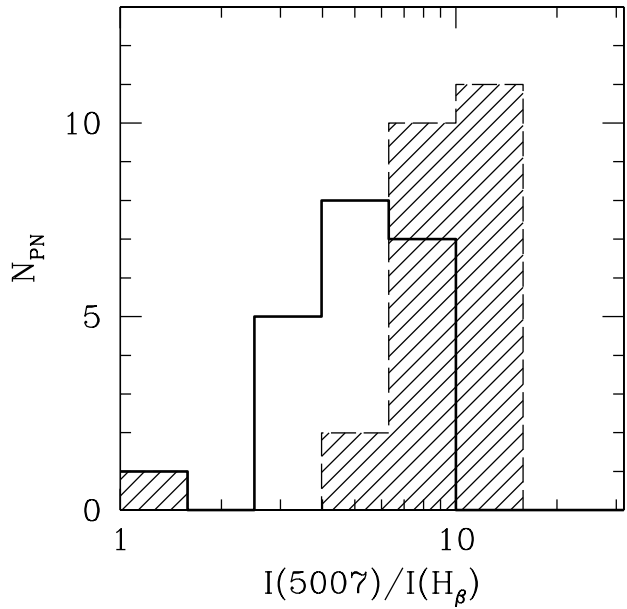}
  \caption{Distribution of the \oiii $\lambda$5007 over \hb intensity ratios
in the SMC (thick histogram) and LMC (shaded histogram) PNs, from Stanghellini et al. 2003a.}\label{fig3}
\end{figure}

The \oiii/\hb emissivity ratio is physically scaled linearly with 
the O/H abundance and the fractional ionization of O$\rm^{++}$. Also it 
depends exponentially on the local electron excitation temperature, 
T$_e$(O$\rm^{++}$) since electron collisions on the high-energy tail of 
the free energy distribution excite the transition.  Of course, 
T$_e$(O$\rm^{++})$ depends on O/H and O$\rm^{++}$/O as well.  So 
interpreting the differences between the \oiii/\hb ratios of the 
SMC and the LMC is best done using ionization models. 

Our Cloudy (Ferland 1996) models explore the major line emission
in a set of Galactic, LMC, and SMC models with same gas density (1000 cm$^{-3}$)
and different metallicities, adequately chosen to represent 
the {\it average} nebula in each studied galaxy, as explained in Stanghellini et
al. (2003a).
The stellar ionizing spectrum 
is assumed to be a blackbody with temperatures and luminosities from the 
H-burning evolutionary tracks for the appropriate galaxian population
by Vassiliadis and Wood (1994).

 \begin{figure}
\includegraphics[height=8truecm]{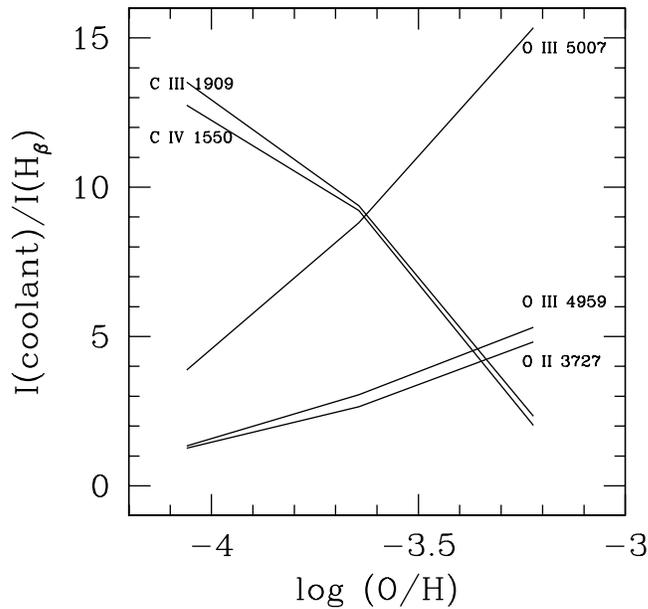}
  \caption{Intensity ratios of the major PN coolants over \hb, versus oxygen
abundances (from Stanghellini et al. 2003a).
  }\label{fig4}
\end{figure}

In Figure 4 we show the line intensity relative to \hb
for the major coolants in the SMC, LMC, and Galactic PNs, versus the oxygen abundance,
as derived from our simplified Cloudy models. 
While we have calculated the evolution of these intensity ratios following the
evolution of the CS from
the early post-AGB phase to the white dwarf stage, 
we only plot here the flux ratios corresponding to the models
with the highest temperature, for each PN
composition.  In general, our target selection tends to favor targets with hottest 
CSs: T$_{eff} \ge 50,000$ K both in the SMC and in the LMC, thus
the set of high-temperature models is the most adequate to reproduce the 
observations for LMC and SMC PNs.

The cooling processes that determine T$_e$(O$\rm^{++})$ in the SMC, LMC 
and Galactic PNs are noteworthy.  In the Galaxy the primary coolants of 
PNs with hot CSs are the optical forbidden lines of \oiii
$\lambda$5007 and other lines of O$\rm^+$ and O$\rm^{++}$.  However, 
in environments in which O/H is as low as in the SMC, the primary 
coolants may become ultraviolet intercombination lines of 
C$\rm^+$ and C$\rm^{++}$.  
The simple models described here seem to reproduce very well the optical
flux ratios of PNs in the Magellanic Clouds.
It will be interesting to confirm these 
predictions with future UV observations.

\subsection{Chemical analysis of LMC PNs}

Studies of Galactic PNs have shown that the bipolar and, more
in general, asymmetric PNs
present lower carbon abundance (and higher nitrogen abundance) than their symmetric (round or
elliptical) counterparts (Peimbert 1978; Stanghellini et al. 2002b).
The reason for the carbon underabundance is to be found in late phases
of AGB evolution, when the nuclear burning extends to the bottom of the 
stellar envelope (HBB, see Iben \& Renzini 1983).
The result seem to imply that asymmetric PNs are the progeny of the more massive stars 
(M$>$3\sm) in the PN progenitor range, since only these stars undergo the HBB process
(for the yields of the LMC AGB stars see Van den Hoek \& Groenevegen
1997).
A statistically sound result is not possible in the Galactic environment, since
the asymmetric PNs tend to be located near the Galactic plane, thus their observation
may be severely affected by the Galactic plane reddening.
With our LMC observations we are in the position to test
these findings by circumventing the selection effects. We found that carbon abundance 
is higher than 1$\times 10 ^{-4}$ for all round and elliptical PNs, while it can be
as low as $\approx 3\times 10 ^{-6}$ for the asymmetric PNs (Stanghellini, Shaw,
Balick, \& Blades 2000, all abundances are by number, versus hydrogen), in agreement with the 
trends that were expected form the Galactic results.

We have also studied the abundance-versus-morphology relations for the elements
whose abundance do not dramatically change
during the evolution of the PN progenitors, such as sulfur, argon,
oxygen, and neon (the so-called {\it alpha elements}). In Figure 5 we show the 
relations between neon and,
from the top panel to the lower panel, oxygen, argon, and sulfur abundances. 
The LMC PNs plotted are coded for morphological type, as described in the label.
The large circled symbols represent the {\it average} LMC abundance of the given
elements in the H~II regions. We see that there is morphological discrimination
between the high and low alpha-element abundance PNs.

  \begin{figure}
\includegraphics[height=16truecm]{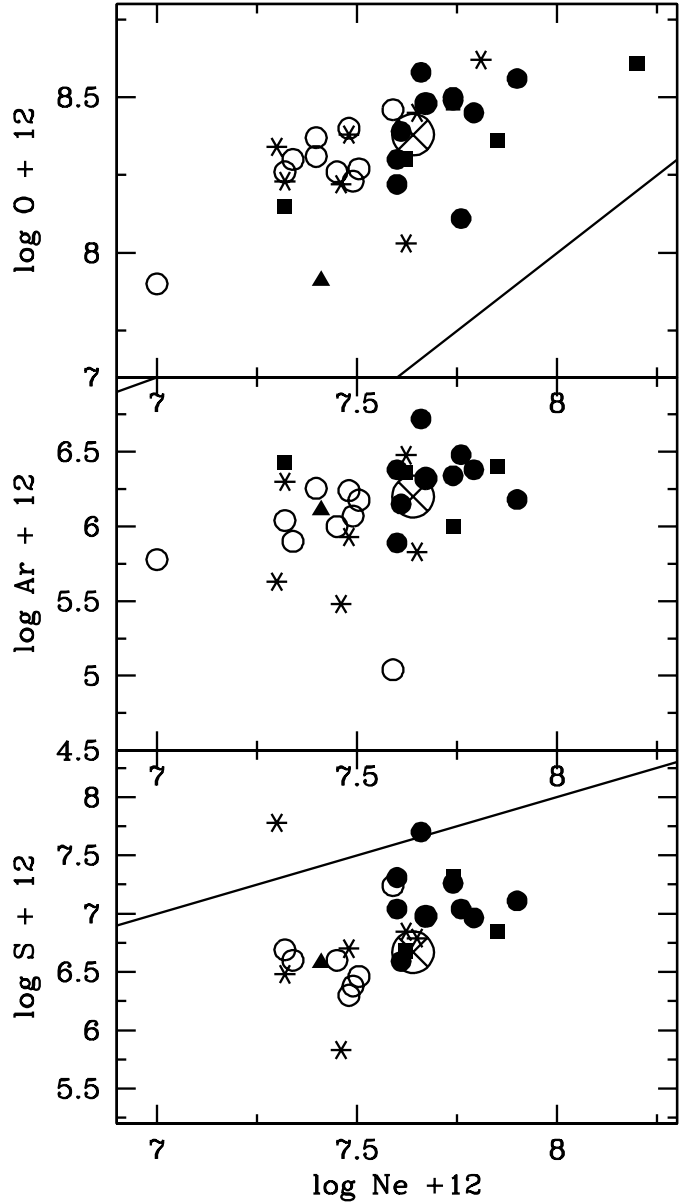}
  \caption{O, Ar, and S vs. Ne abundances of LMC PNs for morphological types round 
({\it open circles}), elliptical ({\it asterisks}), quadrupolar 
({\it filled triangles}), bipolar core ({\it filled circles}), and 
bipolar ({\it filled squares}). The large crossed circle represents
the average for LMC H~II regions (adapted from Stanghellini et al. 2000).
  }\label{fig5}
\end{figure}

\subsection{Central stars of LMC PNs}

Our {\it HST} programs allowed, for the first time, the direct
observation of CSs of extra-Galactic PNs. While
the observations in slitless mode were specially crafted to detect the
nebulae in the most prominent emission lines, the broad-band imagery 
showed the CSs in approximately 60$\%$ of the targets. 

We have measured magnitudes via aperture photometry of the CSs in the
LMC (Villaver, Stanghellini, \& Shaw 2003) and SMC PNs. By means of the Zanstra analysis, and using the
most recent, uniformly selected He II $\lambda$4686 fluxes, we have
determined the (H I and He II) Zantsra stellar temperatures. Finally,
by placing the stars on the log L - log T$_{\rm eff}$ plane, and comparing their loci 
with the theoretical evolutionary post-AGB tracks by Vassiliadis \& Wood
(1994) we have determine the stellar mass.

The median mass of LMC and SMC CSs, respectively 0.63 and 0.59
\sm, do not differ very much,
indicating that the possible variance
with metallicity of the factors at play (initial mass function, 
star formation rate and history, and mass-loss rate) 
probably average each other out (Villaver, Stanghellini, \& Shaw, in preparation).

The total number of CSs analyzed so far, whose masses are determined
with confidence, is small (approximately 25 objects total). We will examine
a larger sample of CSs in the Clouds to deepen the analysis of
possible correlations between the core mass and the morphology of PNs
in the future.

Knowing the position of the CSs on the log L - log T$_{\rm eff}$ diagram is essential
not only for the mass determination, but also to constrain the transition time
(time interval between the quenching of the envelope ejection at the AGB tip and the
PN illumination, see Stanghellini \& Renzini 2000).
Stellar evolutionary models for post-AGB stars 
start with {\it ad hoc} models that imply strong (or semi-empirical)
assumptions on the mass loss rate. From an observational viewpoint, the timing of post-AGB
evolution is derived from the dynamic time of the PN. For Galactic PNs
the comparison of the dynamic and evolutionary time-scales 
has the double offset of the (undetermined)
transition time and the distance bias, while for LMC and SMC PNs the
difference between the observed and theoretical evolutionary time
allows the estimate of the transition time.

The transition time is relevant in post-AGB population synthesis, for
example in the studies of the 
PNLF
as a secondary distance-scale indicator, and also in studies 
concerning the 
UV contribution from post-AGB stars in galaxies.
Further constraints on the effective temperature of the CSs
derived from the UV continuum fitting, from our Cycle 10 UV
spectra, will strengthen the results on stellar evolution.

\section{Summary}

Magellanic PNs are ideal probes to study stellar evolution and populations of 
low- and intermediate-mass stars. The use of the 
{\it HST} is fundamental for determining the PN shapes, the radii, and
also to detect the CSs. Furthermore, only with the use of spatially
resolved images one can identify the LMC and SMC PNs unambiguously,
without the accidental inclusion of compact H II regions in the PN samples. 

We have presented the results derived from our {\it HST}
programs, in the background of the important work that has been done in
LMC and SMC PNs previously.

We found that PNs have the same morphological types in the Galaxy, the LMC, and
the SMC. We also found that the distribution of the morphological types
is noticeably different in the SMC and the LMC, and that the LMC seems to be
populated by PNs whose progenitors are, on average, more massive.

We analyzed morphology and chemical composition of a sample of LMC PNs, 
and found that asymmetric (bipolar and bipolar core)
PNs have higher Ne, Ar, and S abundance than symmetric (round and elliptical)
PNs in the LMC, a confirmation that  
they trace more recent stellar populations. 

An empirical relation between the nebular radii and the surface brightness
is found to hold in both SMC and LMC PNs, independent of morphological type.
The relation, once calibrated, will be used to determine the distance scale for
Galactic extended PNs. 

The PN cooling is affected by metallicity, and it seems that the \oiii $\lambda$5007
emission is not always the ideal line to detect bright PNs in all Galaxies,
since the strongest cooling lines in very low metallicity PNs
seem to be the UV \ciii (and \civ) semiforbidden emission.

The observed Magellanic CSs that we discuss here 
constitute the first sizable sample of CS beyond the Milky Way that
has been directly observed. While we found only marginal differences
between the LMC and the SMC median CSs masses of the CSs, we need to enlarge the sample of CS 
whose masses can be reliably measured, given the importance of knowing 
initial-to final-mass relation in different metallicity environments.

\begin{acknowledgements}
The work presented in $\S$4 has been developed in collaboration
with Dick Shaw, Eva Villaver, Bruce Balick, and Chris Blades.  
\end{acknowledgements}

\end{document}